\documentclass[aps,prc,twocolumn,superscriptaddress,showpacs]{revtex4-2}
\usepackage[utf8]{inputenc}
\usepackage{graphicx}
\graphicspath{ {./images/} }
\usepackage{amssymb}
\usepackage{xcolor}
\RequirePackage{amsmath}
\RequirePackage{amsfonts}
\RequirePackage{graphicx}
\usepackage{fontenc}
\usepackage{longtable}
\usepackage{dcolumn}
\usepackage{svrsymbols}
\usepackage{array}
\usepackage{CJKutf8}
\newcolumntype{P}[1]{>{\centering\arraybackslash}p{#1}}
\usepackage[pagebackref=false]{hyperref}	
\hypersetup{
  colorlinks = true,
  linkcolor=blue,   
  citecolor=blue,   
  urlcolor=blue,    
  pdfdisplaydoctitle=true
}

\begin{document}
\title{Direct measurement of Enhanced octupole collectivity in $^{148}_{~66}$Dy}
\begin{CJK*}{UTF8}{gbsn}
\author{P.~Spagnoletti}
\altaffiliation{Corresponding author}
\affiliation{Department of Physics, University of Liverpool, Liverpool L69 7ZE, United Kingdom}
\author{V.~Vedia}
\affiliation{CERN, CH-1211 Geneva, Switzerland}
\author{E.~Y\"uksel} 
\affiliation{School of Mathematics and Physics, University of Surrey, Guildford, GU2 7XH, United Kingdom}
\author{Y.~X.~Yu}
\author{G.~J.~Fu}
\affiliation{School of Physics Science and Engineering, Tongji University, Shanghai 200092, China}
\author{R.~Umashankar}
\affiliation{TRIUMF, 4004 Wesbrook Mall, Vancouver, BC, V6T 2A3, Canada}
\affiliation{Department of Physics and Astronomy, University of British Columbia, Vancouver, British Columbia, Canada V6T 1Z4}
\author{G.~Andreetta}
\affiliation{Istituto Nazionale di Fisica Nucleare, Laboratori Nazionali di Legnaro, Legnaro (PD), Italy}
\affiliation{Dipartimento di Fisica e Astronomia, Università degli Studi di Padova, Padova, Italy}
\author{C.~Andreoiu}
\affiliation{Department of Chemistry, Simon Fraser University, Burnaby, British Columbia V5A 1S6, Canada}
\author{A.A.~Avaa}
\affiliation{TRIUMF, 4004 Wesbrook Mall, Vancouver, BC, V6T 2A3, Canada}
\author{G.C.~Ball}
\affiliation{TRIUMF, 4004 Wesbrook Mall, Vancouver, BC, V6T 2A3, Canada}
\author{V.~Bildstein}
\affiliation{Department of Physics, University of Guelph, Guelph N1G 2W1 Ontario, Canada}
\author{S.~Buck}
\affiliation{Department of Physics, University of Guelph, Guelph N1G 2W1 Ontario, Canada}
\author{E.~Cantacuzene}
\affiliation{Department of Physics, University of Regina, Regina S4S 0A2 Saskatchewan, Canada}

\author{G.~Colombi}
\affiliation{Department of Physics, University of Guelph, Guelph N1G 2W1 Ontario, Canada}
\author{I.~Dillmann}
\affiliation{TRIUMF, 4004 Wesbrook Mall, Vancouver, BC, V6T 2A3, Canada}
\affiliation{Department of Physics and Astronomy, University of Victoria, Victoria, British Columbia, Canada}
\author{L.~P.~Gaffney}
\affiliation{Department of Physics, University of Liverpool, Liverpool L69 7ZE, United Kingdom}
\author{A.B.~Garnsworthy}
\affiliation{TRIUMF, 4004 Wesbrook Mall, Vancouver, BC, V6T 2A3, Canada}
\affiliation{Department of Physics and Astronomy, University of Victoria, Victoria, British Columbia, Canada}
\author{P.E.~Garrett}
\affiliation{Department of Physics, University of Guelph, Guelph N1G 2W1 Ontario, Canada}
\affiliation{Department of Physics and Astronomy, University of the Western Cape, P/B X17, Bellville, ZA-7535, South Africa}
\author{G.F.~Grinyer}
\affiliation{Department of Physics, University of Regina, Regina S4S 0A2 Saskatchewan, Canada}
\author{G.~Hackman}
\affiliation{TRIUMF, 4004 Wesbrook Mall, Vancouver, BC, V6T 2A3, Canada}
\affiliation{Department of Chemistry, Simon Fraser University, Burnaby, British Columbia V5A 1S6, Canada}
\author{J.~Liu}
\affiliation{TRIUMF, 4004 Wesbrook Mall, Vancouver, BC, V6T 2A3, Canada}
\affiliation{Department of Physics and Astronomy, University of Victoria, Victoria, British Columbia, Canada}
\author{A.~Ludlam}
\affiliation{TRIUMF, 4004 Wesbrook Mall, Vancouver, BC, V6T 2A3, Canada}
\author{L.L.~Luperi}
\affiliation{IRFU/DPhN, CEA Saclay, Universit\'e Paris-Saclay, 91191 Gif-sur-Yvette, France}
\author{Madhu}
\affiliation{Department of Chemistry, Simon Fraser University, Burnaby, British Columbia V5A 1S6, Canada}
\author{T.~La~Marca}
\affiliation{INFN Sezione di Firenze, IT-50019 Firenze, Italy}
\author{D.~Movilla-Quintero}
\affiliation{Instituto de Estructura de la Materia, CSIC, Madrid, ES-28006, Spain}
\author{S.~Murillo~Morales}
\affiliation{TRIUMF, 4004 Wesbrook Mall, Vancouver, BC, V6T 2A3, Canada}
\author{R.~Nicolás del Álamo}
\affiliation{Istituto Nazionale di Fisica Nucleare, Sezione di Padova, Padova, Italy}
\affiliation{Dipartimento di Fisica e Astronomia, Università degli Studi di Padova, Padova, Italy}
\author{M.M.~Rajabali}
\affiliation{Physics Department, Tennessee Technological University, Cookeville, Tennessee 38505, USA}
\author{P.H.~Regan}
\affiliation{School of Mathematics and Physics, University of Surrey, Guildford, GU2 7XH, United Kingdom}
\affiliation{National Physical Laboratory, Hampton Road, Teddington, Middlesex, TW11 0LW, United Kingdom}
\author{M.~Rocchini}
\affiliation{INFN Sezione di Firenze, IT-50019 Firenze, Italy}
\author{M.~Scheck}
\affiliation{RR, Forschungsreaktor, Helmholtz Zentrum hereon, 21502 Geesthacht, Germany}
\author{J.~Seger}
\affiliation{Physics Department, Tennessee Technological University, Cookeville, Tennessee 38505, USA}
\author{S.~Sekal}
\affiliation{Physics Department, Tennessee Technological University, Cookeville, Tennessee 38505, USA}
\author{D.~Stramaccioni}
\affiliation{Istituto Nazionale di Fisica Nucleare, Laboratori Nazionali di Legnaro, Legnaro (PD), Italy}
\affiliation{Dipartimento di Fisica e Astronomia, Università degli Studi di Padova, Padova, Italy}
\author{C.E.~Svensson}
\affiliation{TRIUMF, 4004 Wesbrook Mall, Vancouver, BC, V6T 2A3, Canada}
\affiliation{Department of Physics, University of Guelph, Guelph N1G 2W1 Ontario, Canada}
\author{A.~Tsantiri}
\affiliation{Department of Physics, University of Regina, Regina S4S 0A2 Saskatchewan, Canada}
\author{J.~Williams}
\affiliation{TRIUMF, 4004 Wesbrook Mall, Vancouver, BC, V6T 2A3, Canada}
\author{F.~Wu~(吴桐安)}
\affiliation{Department of Chemistry, Simon Fraser University, Burnaby, British Columbia V5A 1S6, Canada}

\begin{abstract}
Excited states in $^{148}_{~66}$Dy were populated via $\beta^+/EC$ decay of $^{148m}$Ho using the GRIFFIN spectrometer at the TRIUMF ISAC-I facility.
A combined measurement of the mean lifetime of the $3_1^-$ level using the Generalized Centroid Difference (GCD) method and branching fraction of the $3_1^-\rightarrow0_1^+$ $\gamma$-ray decay has been performed. 
From these results, an enhanced electric octupole $B(E3;3_1^-\rightarrow0_1^+)$ transition strength of 46(3)~W.u. has been determined in $^{148}_{~66}$Dy.
This is the largest measured value across the closed neutron shell at $N=82$ and provides direct evidence of enhanced octupole collectivity beyond $Z=64$.
The evolution of the $B(E3; 3^-_1 \rightarrow 0^+_1)$ strength along the $N=82$ isotonic chain is compared with quasiparticle random-phase approximation (QRPA) calculations using the SkI3 and SkM$^*$ Skyrme energy-density functionals, as well as with large-scale shell-model (SM) calculations.
This result extends the boundaries of enhanced octupole collectivity far from the so-called `octupole magic numbers' $Z=56$ and $N=88$.

\end{abstract}
\maketitle
\end{CJK*}
The breaking of reflection symmetry in the intrinsic frame can arise from long-range octupole-octupole correlations leading to the existence of ``pear''-shaped nuclei, which possess either static or dynamic octupole deformation~\cite{Butler1996}.
These correlations have a microscopic origin and are strongest when pairs of orbitals which differ in total ($j$) and orbital ($\ell$) angular momentum by 3$\hbar$ lie close to the Fermi surface for both protons and neutrons~\cite{Chen2021,Scheck_2025}. 
Such states approach each other when $Z,N\approx34,56,88$ and $N\approx134$ which leads to specific regions of the nuclear chart, where octupole correlations are strongest.
Considerable focus from theory and experiment has been directed towards the actinides ($Z\sim88$, $N\sim134$) and the lanthanides ($Z\sim56$, $N\sim88$)~\cite{Butler2016_Review}. 

Electromagnetic transition rates between the ground state with spin and parity $0^+$ and the first excited $3^{-}$ are a robust test of octupole collectivity in nuclei with an even number of both protons and neutrons.
However, electric-octupole $(E3)$ transition strengths, $B(E3)$, are challenging to access experimentally as $E3$ transitions typically do not compete when either $E1$ or $E2$ transitions are allowed to depopulate the considered level.
Currently, only the $^{222,224,226}$Ra~\cite{Gaffney2013,Wollersheim1993,Butler2020} ($Z=88$) nuclei which, exhibit enhanced $B(E3)$ values consistent with the rotational model, can be reliably considered to possess stable octupole deformation.
Despite their close proximity, $^{228}$Ra~\cite{Butler2020} and $^{220-226}$Rn~\cite{Butler2019,Butler2020b,Spagnoletti2022} display experimental signatures indicating dynamic octupole shapes, meaning small changes in proton and neutron numbers lead to consequential differences in nuclear shape.

Within the lanthanide region ($Z\sim56,~N\sim88$) of the nuclear chart, considerable interest was centered around enhanced $B(E3;3_1^-\rightarrow0_1^+)=48^{+25}_{-34}$~W.u.~and $48^{+21}_{-29}$~W.u.~measured in $^{144}$Ba~\cite{Bucher2016} and $^{146}$Ba~\cite{Bucher2017}, respectively, and the suggestion that these nuclei possess stable octupole deformation, albeit with large uncertainties.
However, new results with drastically improved precision in $^{142,144}$Ba~\cite{144Ba} present no strong enhancement of $B(E3)$ strength which are more consistent with beyond mean-field (BMF) theoretical approaches~\cite{BMF1,BMF2,BMF3,BMF4} and provide convincing evidence that the neutron-rich Ba isotopes are octupole vibrational in nature.
It remains unclear whether any nuclei in this region possess stable octupole deformation where the neighboring odd-mass systems are the favored candidates~\cite{EDM-Candidates,Schiff-Moment-Flambaum} to investigate $CP$-violations with the search for permanent atomic electric-dipole moments (EDM)~\cite{Chupp2019}.
Odd-mass systems with stable octupole deformation provide an enhancement of the nuclear Schiff moment, which induces the atomic EDM, but this enhancement can occur to a lesser extent in octupole vibrational systems~\cite{Dalton2023} such as $^{153}$Eu $(Z=63)$~\cite{Schiff-Moment-Sushkov}. 
What is clear from experimental data, is a direct relationship of increasing $B(E3)$ strength with increasing atomic number from $Z=50$ to 64 within the boundaries of $N=82-88$, where the largest measured $B(E3;3_1^-\rightarrow0_1^+)$=45(5)~W.u.~is in $^{150}$Gd~\cite{Pascu-150Gd} $(Z=64,N=86)$.
However, it is unknown if this relationship continues beyond $Z=64$ to heavier elements or if the Gd isotopes possess the strongest $E3$ strength in the region.
This is further complicated by large reductions in $B(E3)$ strength from $N=88$ to $90$ observed in the Nd, Sm and Gd isotopes~\cite{KIBEDI_BE3}, which coincides with a quantum shape-phase transition~\cite{shape-phase-transition-casten-prl} and large changes in quadrupole deformation are observed~\cite{Casten2006}.
This reduction could arise as other modes of octupole shape
oscillations that occur in deformed nuclei may lead to a fragmentation of $B(E3)$ strength to other $3^-$ states~\cite{Butler-framentaion-E3-strength}.

The $N=82$ isotonic chain is within reach of Shell-Model calculations and possess more measured $B(E3;3_1^-\rightarrow0_1^+)$ values than for any other neutron number, providing a robust testing ground to compare theory and experiment.
This chain displays increasing $B(E3)$ strength from the doubly-magic $^{132}$Sn to the quasi-doubly-magic $^{146}$Gd which is formed by the closure of the $\pi d_{5/2}$ subshell at $Z=64$~\cite{quasi-magic-146Gd-1,quasi-magic-146Gd-2,PhysRevLett.41.289,PhysRevLett.47.1433}.
The closed neutron shell preserves a near spherical shape for the ground state and the $3_1^-$ state corresponds to an octupole phonon coupled to the ground state.
This increase in $B(E3)$ strength is reflected in a lowering of the $3_1^-$ state excitation energy which reaches a minimum at $^{146}$Gd, becoming the first excited state similar to $^{208}$Pb, and then starts increasing for heavier elements.
For $^{148}$Dy, the $3_1^-$ state is positioned at 1687~keV just above the $2_1^+$ state at 1677~keV.
These excitation energies, combined with the natural hindrance of electric-dipole ($E1$) transitions~\cite{E1-Hindered,ANDREJTSCHEFF2001239}, make $^{148}$Dy one of the rare easily accessible $3_1^-\rightarrow0_1^+$ $E3$ decays. 
 
In this letter, we report a measurement of the mean lifetime of $3_1^-$ level using fast-timing methods and the absolute $\gamma$-ray branching fraction of the $E3$ transition to the ground-state to provide a measure of the $B(E3;3_1^-\rightarrow0_1^+)$ strength in $^{148}$Dy $(Z=66)$ with high precision.

Excited states in $^{148}$Dy were populated via $\beta^+/EC$ decay of the $I^{\pi}=5^-$ isomer of $^{148m}$Ho ($T_{1/2}=9.60(13)$~s, $Q_{\beta^+} > 8.85(80)$~MeV) produced by bombarding a Ta target with 480-MeV protons at the TRIUMF ISAC-1 facility~\cite{Dilling2014}.
The Ho atoms were ionized using the ion-guide laser ion source~\cite{IGLIS}, which was exploited for its suppression capabilities of easily-surface-ionized isobaric contaminants with dominant production cross sections, then mass separated to produce a near pure beam of $^{148}$Ho ions.
A 40~keV beam, with an intensity of $\approx$5000 pps, was implanted onto a Mylar tape system situated at the center of the GRIFFIN spectrometer~\cite{GRIFFIN_NIM}.
GRIFFIN was equipped with fifteen HPGe clover detectors and eight LaBr$_3$(Ce) fast-scintillator detectors both with active anti-Compton BGO shielding, the zero-degree plastic scintillator for $\beta$~particles and PACES, an array of five Si(Li) detectors, for conversion~electrons.
Decay data were collected in cycles to maximize the total decays of $^{148m}$Ho and minimize the activity of the decay chain of $^{148}$Dy ($T_{1/2}=3.2(2)$~m) using an implantation period of 45~s and a further 20~s with no implantation.
The tape was moved after each cycle, positioning the implantation point behind a lead-shielded box to minimize background from the decay chain. 

The data were collected using the GRIFFIN digital data acquisition system operating in a triggerless mode.
Additionally, the anode signals from the LaBr$_3$(Ce) PMTs were input into Ortec 935 quad Constant Fraction Discriminator (CFD) modules and the subsequent output signals of the individual CFDs were fed into a Lecroy 429A Quad Mixed Logic Fan-In/Fan-Out modules.
The logic modules provide the start and stop signals to a set of Ortec 566 time-to-amplitude converter (TAC) modules, where the amplitude of the TAC signal corresponds to the time difference between the start and stop signals.
The gains of the TAC modules were calibrated using an Ortec 462 time-calibrator module with an average of 3.3~ps/channel.
The prompt TAC signals of all possible detector combinations were aligned offline using the 1173-keV and 1333-keV $\gamma$-ray coincidence emitted from a $^{60}$Co calibration source. 
Further details on the electronic fast-timing technique applied to GRIFFIN are available in Refs.~\cite{GRIFFIN_NIM,Bruno-Griffin-fast-timing}.

\begin{figure}[!t]
\includegraphics[width=\columnwidth]{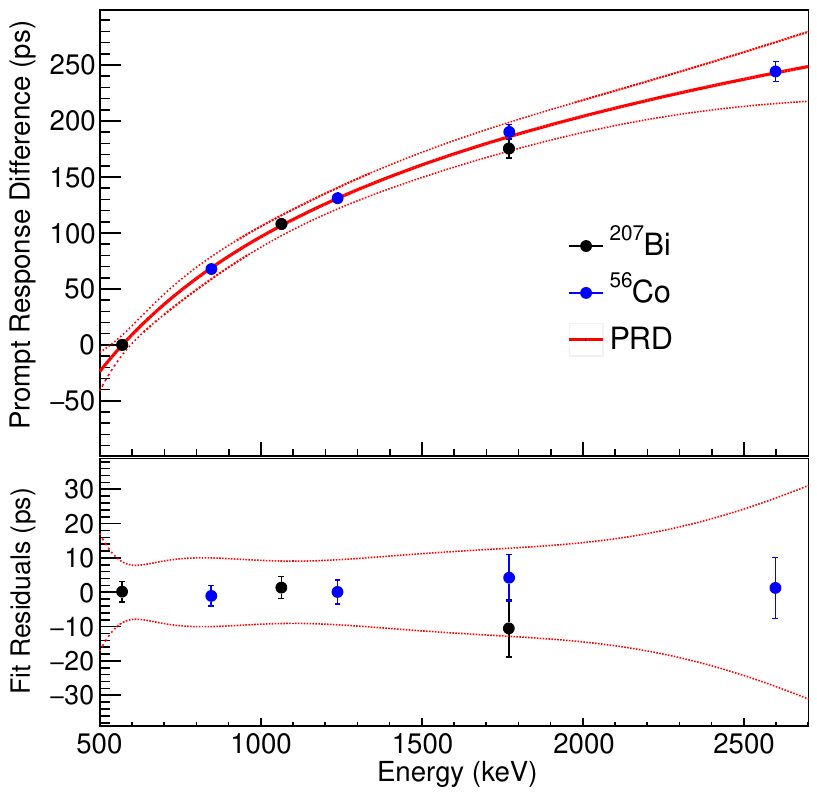}
\caption{Top: Prompt response difference (PRD) measured with calibration sources $^{207}$Bi and $^{56}$Co. Bottom: PRD fit residuals color separated for each reference. A $3\sigma$ confidence interval (CI) from the fit is adopted.}
\label{prd-fit}
\end{figure}

%
The mean lifetime of the $3_1^-$ level was obtained using the coincidence $\gamma$-ray decay fast-timing technique with the Generalized Centroid Difference (GCD) method~\cite{Regis-GCD,REGIS2025104152}.
The method employs a TAC to measure the time difference between two $\gamma$-ray transitions emitted in a cascade, that were detected, in this work, by pairs of LaBr$_3$(Ce) fast scintillator detectors.
The delayed time distribution corresponds to a TAC being started by a $\gamma$-ray transition, which populates the level of interest and is stopped by a $\gamma$-ray decay from said level.
The centroid of the time distribution is displaced by the mean lifetime, $\tau$, of the level.
The reversal between the transitions, which start and stop the TAC, corresponds to the anti-delayed time distribution with the centroid displaced by $-\tau$.
When the time distributions are free from time-correlated background, the lifetime of the level can be related to the difference between the centroids of the delayed and anti-delayed distributions, $\Delta C_{\mathrm{FEP}}$, by the equation
\begin{equation}
\Delta C_{\mathrm{FEP}} = 2\tau + PRD(E_{\mathrm{feeder}},E_{\mathrm{decay}}),
\label{lifetime}
\end{equation}
where FEP are full-energy peak events and the Prompt Response Difference, $PRD(E_{\mathrm{feeder}},E_{\mathrm{decay}})$ is the energy-dependent mean time-walk of the system.
The PRD was determined by measuring the centroid difference of $\gamma-\gamma$ cascades with known lifetimes using $^{207}$Bi and $^{56}$Co calibration sources.
The results are presented in Fig.~\ref{prd-fit}.

\begin{figure}[t]
\includegraphics[width=\columnwidth]{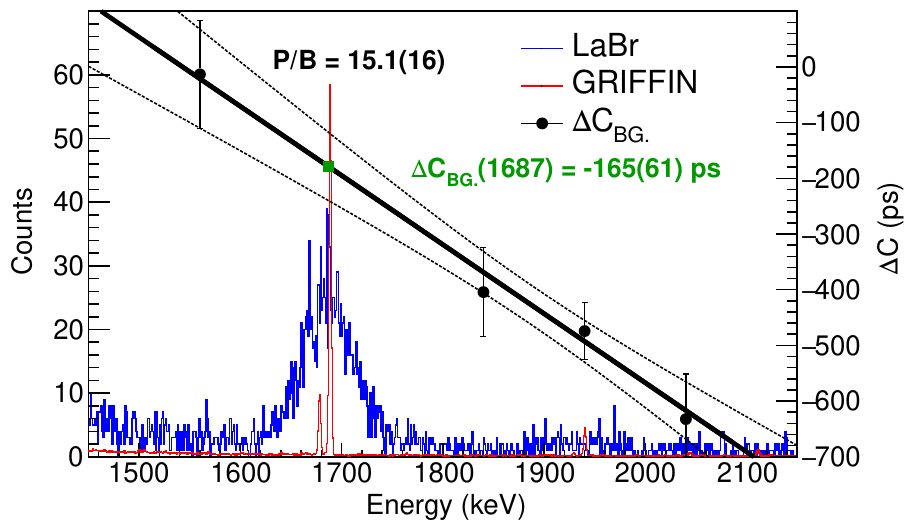}
\includegraphics[width=\columnwidth]{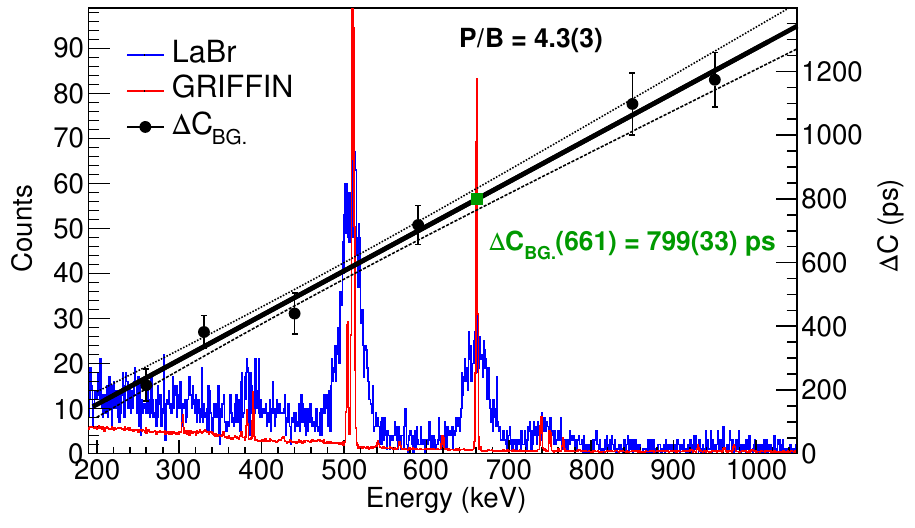}
\caption{Energy spectra of events in the LaBr$_3$(Ce) (blue) and HPGe GRIFFIN (red) detectors in coincidence with the 660.8-keV (top) and 1687-keV (bottom) $\gamma$-ray transitions detected with the LaBr$_3$(Ce) detectors. 
The interpolation of the time responses of the Compton background $\Delta C_{BG.}$ are included.}
\label{labr-spectra}
\end{figure}

\begin{figure}[t]
\includegraphics[width=\columnwidth]{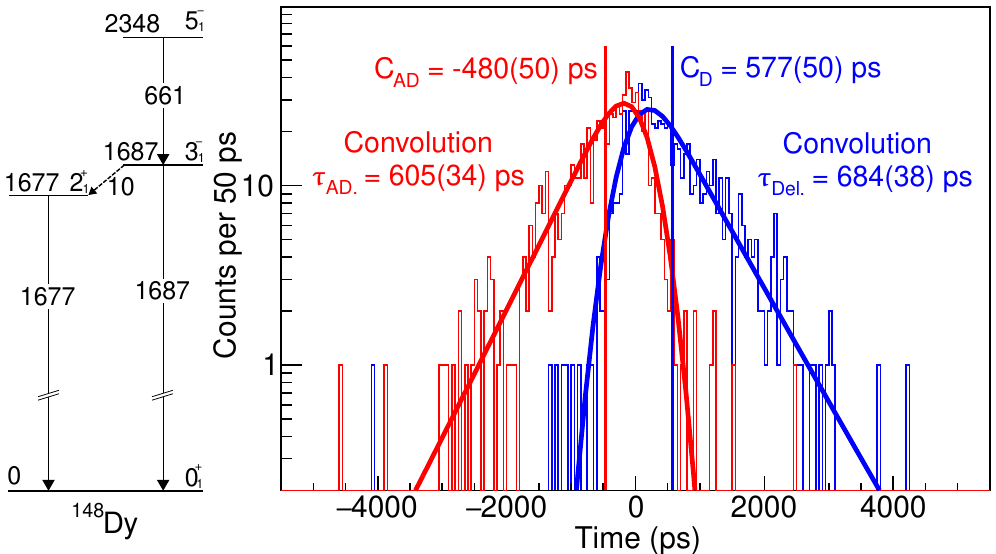}
\caption{Left: Partial level scheme of $^{148}$Dy showing the feeder and decay transitions for the $3_1^-$ level. Level and transition energies are given in keV. Right: The experimental delayed (blue) and anti-delayed (red) time distributions of the $3^-$ state of $^{148}$Dy using the 661-(1677,1687) $\gamma$-$\gamma$ cascade. The solid lines correspond to a fit of the distributions using the convolution method~\cite{REGIS2025104152}.
}
\label{tac-spectrum}
\end{figure}

The $\beta^+/EC$ decay of $^{148m}$Ho strongly populates the $5_1^-$ level in $^{148}$Dy both directly from the decay and indirectly via higher-lying levels populated from $\beta^+/EC$ decay.
The $5_1^-$ level then populates the $3_1^-$ level via a 660.8-keV $\gamma$-ray transition and is employed as the feeding transition in our analysis of the timing spectra.
The projection of $\gamma$-ray energies from the $\gamma$-$\gamma$-$t$ coincidence events with a LaBr$_3$(Ce) gate on the 660.8-keV feeding transition is shown in Fig.~\ref{labr-spectra} (top) and includes a spectrum of coincidence events in the GRIFFIN HPGe detectors produced with the same LaBr$_3$(Ce) energy gate.
The GRIFFIN coincidence spectrum, reveals that both the $3_1^-\rightarrow0_1^+$ 1687.5-keV and $2_1^+\rightarrow0_1^+$ 1677.5-keV $\gamma$-ray transitions are present in the LaBr$_3$(Ce) decay energy gate.
The $2_1^+$ level is populated via a 10-keV $E1$ transition from the decay of the $3_1^-$ level, as shown in the partial level scheme in Fig.~\ref{tac-spectrum}, and therefore a delayed time distribution generated via $5_1^-\rightarrow3_1^-~-~2_1^+\rightarrow0_1^+$ (660.8-1677.5) $\gamma-\gamma$ coincidences would correspond to the effective lifetime of the cascade i.e. $\tau_{\mathrm{eff}}~=~\tau(3_1^-)~+~\tau(2_1^+)$. 
However, the neighboring $N=82$ isotones $^{138}$Ba, $^{140}$Ce, $^{142}$Nd and $^{144}$Sm all possess measured $B(E2;2_1^+\rightarrow0_1^+)$ values greater than 10 W.u.~\cite{BE2_comp}~and therefore, using a conservative assumption that, in $^{148}$Dy, $B(E2;2_1^+\rightarrow0_1^+)$~$\ge$~2~W.u., the mean lifetime of the $2_1^+$ level would be $<$~1~ps and consequently, would have a negligible contribution to the mean lifetime measurement of the $3_1^-$ level. 
The bottom of Fig.~\ref{labr-spectra} presents the reversed energy gate for the 1677.5-keV and 1687.5-keV $\gamma$-ray transitions emitted following the decay of the $3_1^-$ level. 

The delayed time distribution, presented in Fig.~\ref{tac-spectrum}, corresponds to a TAC started by a 660.8-keV feeding transition and stopped by either the 1687- or 1677-keV decay transitions. 
The anti-delayed time distributions correspond to the reversal of the start and stop energy conditions.
From the timing spectra, we obtain a measured centroid difference $\Delta C_{\mathrm{exp}}=1057(71)$~ps between the delayed and anti-delayed time distributions.
However, the measured centroid difference possesses contributions from the Compton background which produces a non-negligible time-correlated background and must be accounted for to obtain the true centroid difference.

\begin{figure}[t]
\includegraphics[width=\columnwidth]{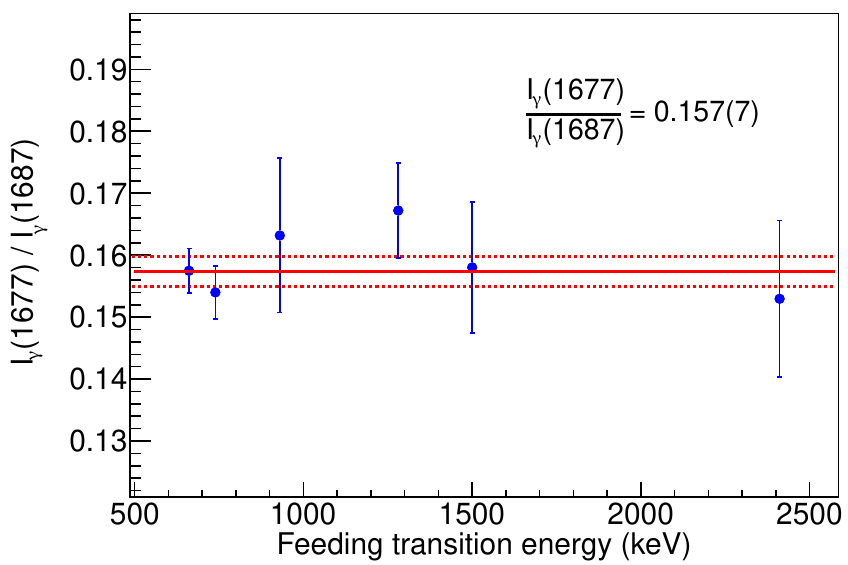}
\caption{Ratio of $2^+ \rightarrow 0^+$ relative to $3^- \rightarrow 0^+$ $\gamma$-ray transition intensities measured in coincidences with higher-lying transitions which directly populated the $3^-$ level. The x-axis represents the energy of the $\gamma$-ray transition feeding the $3^-$ level.}
\label{branching-ratio}
\end{figure}

In this work, we employ the well-established analytical background time correction~\cite{Bk-Corr-1,Bk-Corr-2,Bk-Corr-3,Bk-Corr-4} to reliably remove time contributions from the Compton background.
Here, the experimental centroid difference $\Delta C_{\mathrm{exp}}$ is corrected using the equation:
\begin{equation}
    \Delta C_{\mathrm{FEP}}=\Delta C_{\mathrm{exp}} + \frac{1}{2}(t_{\mathrm{cor}}(E_{\mathrm{feeder}}) +t _{\mathrm{cor}}(E_{\mathrm{decay}})).
\end{equation}
The time corrections, $t_{\mathrm{cor}}$, are defined as
\begin{equation}
    t_{\mathrm{cor}} = \frac{\Delta C_{\mathrm{exp}} - \Delta C_{\mathrm{BG}} }{P/B}, 
    \label{time-corr}
\end{equation}
where $\Delta C_{\mathrm{BG}}$ is the centroid difference of the background time distribution at the considered $\gamma$-ray energy and $P/B$ is the corresponding peak-to-background ratio.
The values of $\Delta C_{\mathrm{BG}}$ are not measured directly but are interpolated from measurements of background time distributions at different energies around each FEP.
The background time responses contributing to both the decay and feeder are presented in the top and bottom of Fig.~\ref{labr-spectra}, respectively and includes the $P/B$ values for each transition.
For the 1687-keV transition we obtain a peak-to-background ratio $P/B$=15.1(16) and $\Delta C_{\mathrm{BG}}(1687)$=-165(61)~ps, while for the 661-keV transition we have $P/B$=4.3(3) and $\Delta C_{\mathrm{BG}}(661)$=799(33)~ps.
From Eq.~\ref{time-corr}, we obtain timing correction values of $t_{\mathrm{cor}}(1687)$=81(11)~ps and $t_{\mathrm{cor}}(661)=60(19)$~ps and a mean value of $\overline{t}_{\mathrm{cor}}=71(21)$~ps.
The mean lifetime of the $3_1^-$ level is derived according to Eq.~\ref{lifetime}:
\begin{equation}
\begin{split}
\tau & = \frac{\Delta C_{\mathrm{exp}} + \overline{t}_{corr} - PRD(661,1687) }{2} \\
 & = \frac{1057(71) + 71(21) - -152(15)}{2} \\
 & = 640(38)\text{~ps},
\end{split}
\end{equation}
where $PRD(661,1687) = PRD(661) - PRD(1687)=-152(15)$~ps, is the prompt response difference of the 661-1687 $\gamma$-$\gamma$ coincidence and was derived from the calibrated PRD curve presented in Fig.~\ref{prd-fit}.
The lifetime, $\tau=640(38)$~ps, extracted from the centroid difference method. 
This is in good agreement with the values obtained via the convolution method for the delayed $\tau_D=684(38)$~ps and anti-delayed $\tau_{AD}=605(34)$~ps time distributions, which have a weighted mean $\overline{\tau}~=~640(25)$~ps and provides additional confidence in our results.
The reduced electric-octupole transition rate is given by the Eq.~\cite{KIBEDI_BE3}:
\begin{equation}
    B(E3)\uparrow = \frac{12260}{E_{\gamma}^{7} \tau_{E3}}e^2b^3,
    \label{be3-eqaution}
\end{equation}
where $E_{\gamma}$ is the $\gamma$-ray energy in MeV and $\tau_{E3}$ is the partial mean lifetime of the $E3$ $\gamma$-ray transition in picoseconds.
The partial lifetime is defined as the mean lifetime normalized by the $E3$ ground-state branching fraction, $I_{\gamma,rel.}(3_1^-\rightarrow0_1^+)$, which is given by the equation:
\begin{equation}
    I_{\gamma,rel.}(3_1^-\rightarrow0_1^+) = \frac{I_{\gamma}^{E3}}{I^{E3}_{\gamma}\cdot(1+\alpha_{E3}) + I^{E1}_{\gamma}\cdot(1+\alpha_{E1})},
\end{equation}
where $I_\gamma^{E3}$ and $I_\gamma^{E1}$ are the intensities of the $3_1^-\rightarrow0_1^+$ and $3_1^-\rightarrow2_1^+$ transitions and $\alpha_{E3,E1}$ are their internal conversion coefficients. 
The competing 10-keV $E1$ transition depopulating the $3_1^-$ state was not observed in this work due to the low detection efficiency at this energy and the high probability that the decay occurs via internally converted electrons with $\alpha_{E1}=27$~\cite{DataSheetsA148}.
We therefore exploit the subsequent 1677-keV $E2$ transition to determine the $E3$ branching fraction.
For this, we measured the intensities of the 1677- and 1687-keV transitions observed in coincidence with $\gamma$-ray transitions that directly populate the $3_1^-$ level and the results are presented in Fig.~\ref{branching-ratio}.
Summing corrections were performed using the $180^{\circ}$ $\gamma$-$\gamma$ coincidence method~\cite{GRIFFIN_NIM} and their contributions were found to be negligible.
After correcting with the internal conversion coefficients~\cite{BRICC} of both transitions we obtain a ground-state branching fraction $I_{\gamma,rel.}(3_1^-\rightarrow0_1^+) = 0.862(11)$, which agrees well with the approximate value of 0.825 derived from Ref.~\cite{DataSheetsA148}.
From our measurements, we extract a partial lifetime $\tau_{E3}=742(45)$~ps.
Finally, for $^{148}$Dy, we obtain from Eq.~\ref{be3-eqaution} a reduced electric-octupole transition rate $B(E3;3_1^-\rightarrow0_1^+)=46(3)$~W.u..
This value, is the largest in the $N=82$ chain and one of the largest within the lanthanide region. 
Additionally, we obtain $B(E1;3_1^-\rightarrow2_1^+)=2.7(3)\times10^{-3}$~W.u.~from the $E1$ $\gamma$-ray branching fraction $I_{\gamma,rel.}(3_1^-\rightarrow2_1^+)=0.0049(4)$. 

\begin{figure}[!t]
\includegraphics[width=\columnwidth]{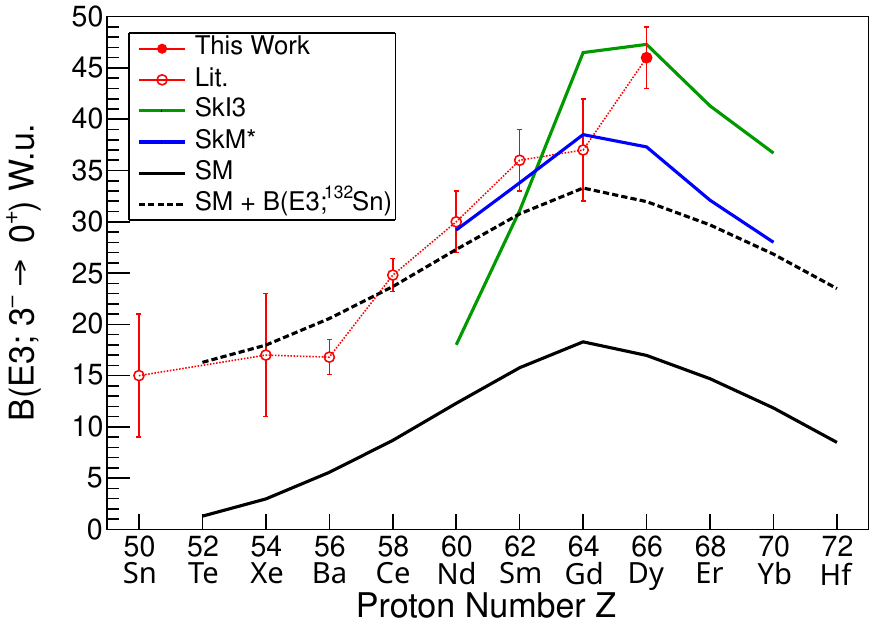}
\caption{Experimental single-particle $B(E3;3_1^-\rightarrow0_1^+)$ values as a function of atomic number from Refs.~\cite{132Sn-CoulEx,KIBEDI_BE3,DataSheetsA146} are presented as open red circles and this work as filled red circles.
The QRPA calculations performed with the SkM* and SkI3 interactions are presented as blue and green lines, respectively.
The solid black line represents the shell-model calculations and the dashed black line shows the same quantity shifted by the experimental $B(E3;3_1^-\rightarrow0_1^+)=15(6)$ W.u.~of the doubly-magic $^{132}$Sn nucleus. }
\label{BE3-rates}
\end{figure}

The evolution of $B(E3;3_1^-\rightarrow0_1^+)$ strength for the $N=82$ chain is presented in Fig.~\ref{BE3-rates}, demonstrating that the increasing $B(E3)$ strength continues beyond $Z=64$.
The experimental data are compared to Quasiparticle Random Phase Approximation (QRPA) and Shell-Model (SM) calculations which, expand upon recent work~\cite{ShellModel-N82} to include $E3$ transition rates. 

The QRPA calculations were performed for the $3^-_1$ states in the even-even $N=82$ isotones within a spherical HF--BCS+QRPA framework
\cite{COLO2013142, colo2021user}. The Skyrme-type SkI3~\cite{Reinhard1995SkI} and SkM*~\cite{QRPA_SkM} interactions were used in the particle--hole channel, while pairing correlations were described with a surface-type pairing interaction. 
Since the calculated $3^-_1$ octupole strength is known to be sensitive to the single-particle structure around the Fermi level and to the associated occupation probabilities, the use of these two interactions also provides a test of this dependence.
The SkM* calculations reproduce the experimental data for $^{142}$Nd, $^{144}$Sm, and $^{146}$Gd very well; however, the present measurement for
$^{148}$Dy shows a much larger enhancement than predicted by SkM*. 
Calculations performed with the SkX interaction in Ref.~\cite{Pascu-150Gd} predict a similar trend to the SkM* calculations in this work, but with consistently lower $B(E3)$ transition strengths. 
In contrast, the SkI3 calculations reproduce the present result for $^{148}$Dy, but predict a similarly large strength in $^{146}$Gd, which is not reflected in the experimental data~\cite{DataSheetsA146}. 
On the lighter side of the chain, the SkI3 calculation predicts a sharp decrease in $B(E3)$ strength towards
$^{142}$Nd. Despite their differences, the QRPA calculations considered here predict the strongest octupole collectivity around $Z=64$--66.

The microscopic structure of the calculated $3^-_1$ states is dominated by the same proton particle--hole component along this part of the chain,
$\pi 1d_{5/2}^{-1}\rightarrow \pi 0h^1_{11/2}$, for both SkI3 and SkM*. 
In $^{148}$Dy, this component accounts for about $84\%$ of the QRPA wave function with SkI3 and about $78\%$ with SkM*. 
Thus, the enhancement from $^{142}$Nd to $^{146}$Gd is not caused by a change in the leading configuration, but by the increasing transition amplitude associated with this octupole-favored proton excitation. 
This reflects the increasing occupation
of the $\pi 1d_{5/2}$ orbital, while the $\pi 0h_{11/2}$ orbital remains largely available as a particle state. 
Around $^{146}$Gd and $^{148}$Dy this mechanism is nearly saturated. 
Beyond this point, the further filling of the $\pi 0h_{11/2}$ orbital reduces the available particle strength, leading to the calculated decrease of the $B(E3)$ strength towards $^{150}$Er and $^{152}$Yb. 
The SkM* calculations show the same underlying mechanism, but with a more fragmented QRPA wave function, including a non-negligible admixture
of the nearby $\pi 1g_{7/2}^{-1}\rightarrow\pi 0h_{11/2}$ component, which does not carry a comparable octupole transition amplitude. 

The SM calculations are thoroughly described in Ref.~\cite{ShellModel-N82}, and therefore, only a brief overview is provided in this text. 
Starting with the doubly-magic nucleus $^{132}$Sn as the inert core and a newly optimized effective proton-proton interaction, the valence protons occupy the $0g_{7/2}$, $1d_{5/2}$, $1d_{3/2}$, $2s_{1/2}$, and $0h_{11/2}$ orbitals of the $Z=50-82$ shell, with neutrons frozen at the $N=82$ closed shell.
The SM calculations performed better in reproducing the excitation energies of the $3_1^-$ levels compared to other interactions~\cite{BROWN2014115,PhysRevC.45.1720,PhysRevC.80.044320}.
However, with the standard SM effective proton charge $e_{\pi}=1.5$, the calculated $B(E3)$ values across the chain are approximately a factor two below the data.
In Ref.~\cite{ShellModel-N82}, this discrepancy is attributed to truncation of the valence space leading to missing contributions from neutron excitations across the $N=82$ shell gap~\cite{ShellGap}.
The full reproduction of $B(E3)$ strength would likely require the valence space to extend across the $Z=50$ and $N=82$ shell closures, successfully performed for the $B(E3)$ in $^{208}$Pb~\cite{Isacker-208Pb}. 
To approximately account for this missing strength, we apply a simple approach of shifting the SM $B(E3)$ values, which arise from protons interacting within the valence space, by the measured $B(E3)$ in $^{132}$Sn~\cite{132Sn-CoulEx}, which is the inert core used in the calculations.
In the doubly-magic nucleus $^{132}_{~50}$Sn, the $B(E3)$ strength arises from the coherent action of $1p1h$ excitations across the proton and neutron shell closures~\cite{Isacker-208Pb}.
Given its simplicity, this approach is remarkably effective for comparing the SM calculations with experimental data.
The large increase in $B(E3)$ strength from $^{146}$Gd to $^{148}$Dy observed in this work is reproduced by neither the SM nor the QRPA calculations, underscoring the need for an accurate description of cross-shell octupole correlations.

In summary, we obtain direct evidence of enhanced octupole collectivity in $^{148}$Dy, which exhibits the largest $B(E3; 3_1^-\rightarrow0_1^+)$ value across the $N=82$ chain.
This is the most precise measurement of enhanced octupole collectivity in the region of the so-called octupole magic numbers $Z=56$, $N=88$. 
This result was obtained from a mean lifetime measurement of the $3_1^-$ level using fast-timing and a precise measurement of the $3_1^-\rightarrow0_1^+$ $\gamma$-ray branching fraction performed with the GRIFFIN spectrometer.
When $E3$ $\gamma$-ray decays are experimentally accessible, our approach is complementary to Coulomb Excitation studies, which is that standard approach for $B(E3)$ measurements~\cite{132Sn-CoulEx,Bucher2016,Bucher2017,Spagnoletti2022,Gaffney2013,Butler2019,Butler2020b,Wollersheim1993,Butler2020,Ibbotson1993}.
This work extends the boundaries of strong enhanced octupole collectivity in the lanthanide region beyond $Z=64$ and provides significant motivation for future studies of the $N=84-88$ Dy isotopes which, based upon the systematics of the $N=82-88$ isotopic chains of Nd, Sm and Gd, and QRPA calculations of $^{148,150,152}$Dy from Ref.~\cite{Pascu-150Gd}, should possess even larger $B(E3; 3_1^-\rightarrow0_1^+)$ strength. 
Finally, with no available $B(E3; 3_1^-\rightarrow0_1^+)$ measurements for the $Z\ge68$ nuclei with $N=82-92$ and an incomplete description of octuple collectivity from theory, the exact boundaries of enhanced octupole collectivity remain an open question.
Therefore, future experimental studies should be directed towards Er and Yb isotopes to obtain a clearer picture of octupole degree of freedom around $Z=56$, $N=88$.

\textit{Acknowledgments}$-$This work would not have been possible without the efforts of the operations and beam delivery staff at TRIUMF for providing the radioactive beam.
This work was supported in part by the Natural Sciences and Engineering Research Council of Canada and by the U.S. Department of Energy, Office of Science, Office of Nuclear Physics, under contract number DE-AC02-06CH11357 and DE-SC0016988339 (TTU).
G.J.F. acknowledges support from the National Natural Science Foundation of China under Grants No. 12322506 and No. 12535009. 
FW acknowledges funding from the Canadian Institute of
Nuclear Physics.
D. M-Q acknowledges support from project PID2022-140162NB-I00, funded by MCIN/AEI/10.13039/501100011033/ERDF, EU and project OTR13111 funded by CSIC-INST/CENTRO.
The infrastructure of GRIFFIN has been funded through contributions from the Canada Foundation for Innovation, TRIUMF, Simon Fraser University, University of Guelph, British Columbia Knowledge Development Fund, and the Ontario Ministry of Research and Innovation. 
TRIUMF receives funding through a contribution agreement through the National Research Council Canada.
This work was supported by the UK Science and Technology Facilities Council through Grant No. ST/Y000242/1, ST/Y000358/1.
\bibliography{sample}

@unpublished{144Ba,
    author = {B.~Jones and L.~P.~Gaffney and M.~Scheck and  P.~Spagnoletti and others},
    title = {Octupole collectivity in $^{142,144}$Ba},
    note = {To be submitted}
}

@article{GRIFFIN_NIM,
title = {The GRIFFIN facility for Decay-Spectroscopy studies at TRIUMF-ISAC},
journal = {Nuclear Instruments and Methods in Physics Research Section A: Accelerators, Spectrometers, Detectors and Associated Equipment},
volume = {918},
pages = {9-29},
year = {2019},
issn = {0168-9002},
doi = {https://doi.org/10.1016/j.nima.2018.11.115},
url = {https://www.sciencedirect.com/science/article/pii/S0168900218317662},
author = {A.B. Garnsworthy and C.E. Svensson and M. Bowry and R. Dunlop and A.D. MacLean and others},
}

@Article{Dilling2014,
author={Dilling, J. and Kr{\"u}cken, R. and Ball, G.},
title={ISAC overview},
journal={Hyperfine Interactions},
year={2014},
month={Jan},
day={01},
volume={225},
number={1},
pages={1-8},
issn={1572-9540},
doi={10.1007/s10751-013-0877-7},
url={https://doi.org/10.1007/s10751-013-0877-7}
}

@article{IGLIS,
    author = {Raeder, Sebastian and Heggen, Henning and Lassen, Jens and Ames, Friedhelm and Bishop, Daryl and Bricault, Pierre and Kunz, Peter and Mjøs, Anders and Teigelhöfer, Andrea},
    title = {An ion guide laser ion source for isobar-suppressed rare isotope beams},
    journal = {Review of Scientific Instruments},
    volume = {85},
    number = {3},
    pages = {033309},
    year = {2014},
    month = {03},
    issn = {0034-6748},
    doi = {10.1063/1.4868496},
    url = {https://doi.org/10.1063/1.4868496},
}

@article{Wollersheim1993,
author = {Wollersheim, H J and Emling, H and Grein, H and Kulessa, R and Simon, R S and Fleischmann, C and de Boer, J and Hauber, E and Lauterbach, C and Schandera, C and Butler, P A and Czosnyka, T},
doi = {10.1016/0375-9474(93)90351-W},
issn = {0375-9474},
journal = {Nucl. Phys. A},
number = {2},
pages = {261--280},
title = {{Coulomb excitation of ^{226}Ra}},
url = {http://www.sciencedirect.com/science/article/B6TVB-473193N-JY/2/bd5fa45e256295d697664fbae3e1d361},
volume = {556},
year = {1993}
}

@article{Butler2020b,
author = {Butler, P. A. and Gaffney, L. P. and Spagnoletti, P. and Konki, J. and Scheck, M. and Smith, J. F. and Abrahams, K. and Bowry, M. and Cederk{\"{a}}ll, J. and Chupp, T. and de Angelis, G. and {De Witte}, H. and Garrett, P. E. and Goldkuhle, A. and Henrich, C. and Illana, A. and Johnston, K. and Joss, D. T. and Keatings, J. M. and Kelly, N. A. and Komorowska, M. and Kr{\"{o}}ll, T. and Lozano, M. and Singh, B. S. Nara and O'Donnell, D. and Ojala, J. and Page, R. D. and Pedersen, L. G. and Raison, C. and Reiter, P. and Rodriguez, J. A. and Rosiak, D. and Rothe, S. and Shneidman, T. M. and Siebeck, B. and Seidlitz, M. and Sinclair, J. and Stryjczyk, M. and {Van Duppen}, P. and Vinals, S. and Virtanen, V. and Warr, N. and Wrzosek-Lipska, K. and Zielinska, M.},
doi = {10.1038/s41467-020-17309-y},
isbn = {4146701910494},
issn = {2041-1723},
journal = {Nat. Commun.},
month = {dec},
number = {1},
pages = {3560},
pmid = {32661232},
publisher = {Springer US},
title = {{Addendum: The observation of vibrating pear-shapes in radon nuclei}},
url = {http://dx.doi.org/10.1038/s41467-020-17309-y http://www.nature.com/articles/s41467-020-17309-y},
volume = {11},
year = {2020}
}

@article{Butler1996,
author = {Butler, P A and Nazarewicz, W},
doi = {10.1103/RevModPhys.68.349},
issn = {0034-6861},
journal = {Rev. Mod. Phys.},
month = {apr},
number = {2},
pages = {349--421},
publisher = {American Physical Society},
title = {{Intrinsic reflection asymmetry in atomic nuclei}},
url = {http://link.aps.org/doi/10.1103/RevModPhys.68.349},
volume = {68},
year = {1996}
}

@article{Butler2016_Review,
author = {Butler, P A},
doi = {10.1088/0954-3899/43/7/073002},
issn = {0954-3899},
journal = {J. Phys. G Nucl. Part. Phys.},
month = {jul},
number = {7},
pages = {073002},
publisher = {IOP Publishing},
title = {{Octupole collectivity in nuclei}},
url = {http://stacks.iop.org/0954-3899/43/i=7/a=073002?key=crossref.be9f7c1e4fdcff6cb3f073cacd8198c7},
volume = {43},
year = {2016}
}

@article{Bucher2017,
author = {Bucher, B. and Zhu, S. and Wu, C. Y. and Janssens, R. V. F. and Bernard, R. N. and Robledo, L. M. and Rodr{\'{i}}guez, T. R. and Cline, D. and Hayes, A. B. and Ayangeakaa, A. D. and Buckner, M. Q. and Campbell, C. M. and Carpenter, M. P. and Clark, J. A. and Crawford, H. L. and David, H. M. and Dickerson, C. and Harker, J. and Hoffman, C. R. and Kay, B. P. and Kondev, F. G. and Lauritsen, T. and Macchiavelli, A. O. and Pardo, R. C. and Savard, G. and Seweryniak, D. and Vondrasek, R.},
doi = {10.1103/PhysRevLett.118.152504},
issn = {0031-9007},
journal = {Phys. Rev. Lett.},
month = {apr},
number = {15},
pages = {152504},
publisher = {American Physical Society},
title = {{Direct Evidence for Octupole Deformation in $^{146}$Ba and the Origin of Large E1 Moment Variations in Reflection-Asymmetric Nuclei}},
url = {http://link.aps.org/doi/10.1103/PhysRevLett.118.152504},
volume = {118},
year = {2017}
}

@article{Ibbotson1993,
author = {Ibbotson, R. W. and White, C. A. and Czosnyka, T and Butler, P. A and Clarkson, N and Cline, D and Cunningham, R. A. and Devlin, M and Helmer, K. G. and Hoare, T. H. and Hughes, J. R. and Jones, G. D. and Kavka, A. E. and Kotlinski, B and Poynter, R. J. and Regan, P. and Vogt, E. G. and Wadsworth, R and Watson, D. L. and Wu, C. Y.},
doi = {10.1103/PhysRevLett.71.1990},
issn = {0031-9007},
journal = {Phys. Rev. Lett.},
month = {sep},
number = {13},
pages = {1990--1993},
publisher = {American Physical Society},
title = {{Octupole collectivity in the ground band of Nd148}},
url = {http://link.aps.org/doi/10.1103/PhysRevLett.71.1990},
volume = {71},
year = {1993}
}

@article{Bucher2016,
author = {Bucher, B. and Zhu, S. and Wu, C. Y. and Janssens, R. V. F. and Cline, D. and Hayes, A. B. and Albers, M. and Ayangeakaa, A. D. and Butler, P. A. and Campbell, C. M. and Carpenter, M. P. and Chiara, C. J. and Clark, J. A. and Crawford, H. L. and Cromaz, M. and David, H. M. and Dickerson, C. and Gregor, E. T. and Harker, J. and Hoffman, C. R. and Kay, B. P. and Kondev, F. G. and Korichi, A. and Lauritsen, T. and Macchiavelli, A. O. and Pardo, R. C. and Richard, A. and Riley, M. A. and Savard, G. and Scheck, M. and Seweryniak, D. and Smith, M. K. and Vondrasek, R. and Wiens, A.},
doi = {10.1103/PhysRevLett.116.112503},
issn = {0031-9007},
journal = {Phys. Rev. Lett.},
month = {mar},
number = {11},
pages = {112503},
publisher = {American Physical Society},
title = {{Direct Evidence of Octupole Deformation in Neutron-Rich Ba 144}},
url = {http://journals.aps.org/prl/abstract/10.1103/PhysRevLett.116.112503},
volume = {116},
year = {2016}
}

@article{Gaffney2013,
title = {Studies of pear-shaped nuclei using accelerated radioactive beams},
author = {Gaffney, L. P. and Butler, P. A. and Scheck, M. and Hayes, A. B. and Wenander, F. and Albers, M. and Bastin, B. and Bauer, C. and Blazhev, A. and Bönig, S. and Bree, N. and Cederkäll, J. and Chupp, T. and Cline, D. and Cocolios, T. E. and Davinson, T. and De Witte, H. and Diriken, J. and  Grahn, T. and Herzan, A. and Huyse, M. and Jenkins, D. G. and Joss, D. T. and Kesteloot, N. and  Konki, J. and Kowalczyk, M. and Kröll, Th. and Kwan, E. and Lutter, R. and Moschner, K. and Napiorkowski, P. and Pakarinen, J. and Pfeiffer, M. and Radeck, D. and Reiter, P. and Reynders, K. and Rigby, S. V. and Robledo, L. M. and Rudigier, M. and Sambi, S. and Seidlitz, M. and Siebeck, B. and Stora, T. and Thoele, P. and Van Duppen, P. and Vermeulen, M. J. and von Schmid, M. and Voulot, D. and Warr, N. and Wimmer, K. and Wrzosek-Lipska, K. and Wu, C. Y. and Zielinska, M.},
journal = {Nature},
volume = {497},
issue = {7448},
pages = {199-204},
numpages = {5},
year = {2013},
month = {May},
publisher = {Nature},
doi = {10.1038/nature12073},
url = {https://doi.org/10.1038/nature12073}
}

@article{Butler2020,
archivePrefix = {arXiv},
arxivId = {2001.09681},
author = {Butler, P A and Gaffney, L P and Spagnoletti, P and Abrahams, K and Bowry, M and Cederk{\"{a}}ll, J and {De Angelis}, G. and {De Witte}, H and Garrett, P E and Goldkuhle, A. and Henrich, C. and Illana, A. and Johnston, K. and Joss, D. T. and Keatings, J. M. and Kelly, N. A. and Komorowska, M. and Konki, J. and Kr{\"{o}}ll, T. and Lozano, M. and Singh, B. S. Nara and O'Donnell, D. and Ojala, J. and Page, R. D. and Pedersen, L G and Raison, C. and Reiter, P. and Rodriguez, J. A. and Rosiak, D. and Rothe, S. and Scheck, M. and Seidlitz, M. and Shneidman, T. M. and Siebeck, B. and Sinclair, J. and Smith, J. F. and Stryjczyk, M. and {Van Duppen}, P. and Vinals, S. and Virtanen, V. and Warr, N. and Wrzosek-Lipska, K. and Zieli{\'{n}}ska, M.},
doi = {10.1103/PhysRevLett.124.042503},
eprint = {2001.09681},
issn = {0031-9007},
journal = {Phys. Rev. Lett.},
month = {jan},
number = {4},
pages = {042503},
title = {{Evolution of Octupole Deformation in Radium Nuclei from Coulomb Excitation of Radioactive $^{222}$Ra and $^{228}$Ra Beams}},
url = {http://arxiv.org/abs/2001.09681 https://journals.aps.org/prl/abstract/10.1103/PhysRevLett.124.042503},
volume = {124},
year = {2020}
}

@article{Butler2019,
author = {Butler, P. A. and Gaffney, L. P. and Spagnoletti, P. and Konki, J. and Scheck, M. and Smith, J. F. and Abrahams, K. and Bowry, M. and Cederk{\"{a}}ll, J. and Chupp, T. and de Angelis, G. and {De Witte}, H. and Garrett, P. E. and Goldkuhle, A. and Henrich, C. and Illana, A. and Johnston, K. and Joss, D. T. and Keatings, J. M. and Kelly, N. A. and Komorowska, M. and Kr{\"{o}}ll, T. and Lozano, M. and {Nara Singh}, B. S. and O'Donnell, D. and Ojala, J. and Page, R. D. and Pedersen, L. G. and Raison, C. and Reiter, P. and Rodriguez, J. A. and Rosiak, D. and Rothe, S. and Shneidman, T. M. and Siebeck, B. and Seidlitz, M. and Sinclair, J. and Stryjczyk, M. and {Van Duppen}, P. and Vinals, S. and Virtanen, V. and Warr, N. and Wrzosek-Lipska, K. and Zielinska, M.},
doi = {10.1038/s41467-019-10494-5},
isbn = {4146701910494},
issn = {2041-1723},
journal = {Nat. Commun.},
month = {jun},
number = {1},
pages = {2473},
publisher = {Springer US},
title = {{The observation of vibrating pear-shapes in radon nuclei}},
url = {http://www.nature.com/articles/s41467-019-10494-5 https://doi.org/10.1038/s41467-019-10494-5},
volume = {10},
year = {2019}
}

@article{Chen2021,
archivePrefix = {arXiv},
arxivId = {2012.06500},
author = {Chen, Mengzhi and Li, Tong and Dobaczewski, Jacek and Nazarewicz, Witold},
doi = {10.1103/PhysRevC.103.034303},
eprint = {2012.06500},
issn = {2469-9985},
journal = {Phys. Rev. C},
month = {mar},
number = {3},
pages = {034303},
title = {{Microscopic origin of reflection-asymmetric nuclear shapes}},
url = {http://arxiv.org/abs/2012.06500 https://link.aps.org/doi/10.1103/PhysRevC.103.034303},
volume = {103},
year = {2021}
}

@article{Spagnoletti2022,
author = {Spagnoletti, P. and Butler, P. A. and Gaffney, L. P. and Abrahams, K. and Bowry, M. and Cederk{\"{a}}ll, J. and Chupp, T. and de Angelis, G. and {De Witte}, H. and Garrett, P. E. and Goldkuhle, A. and Henrich, C. and Illana, A. and Johnston, K. and Joss, D. T. and Keatings, J. M. and Kelly, N. A. and Komorowska, M. and Konki, J. and Kr{\"{o}}ll, T. and Lozano, M. and Singh, B. S. Nara and O'Donnell, D. and Ojala, J. and Page, R. D. and Pedersen, L. G. and Raison, C. and Reiter, P. and Rodriguez, J. A. and Rosiak, D. and Rothe, S. and Scheck, M. and Seidlitz, M. and Shneidman, T. M. and Siebeck, B. and Sinclair, J. and Smith, J. F. and Stryjczyk, M. and {Van Duppen}, P. and Vi{\~{n}}als, S. and Virtanen, V. and Wrzosek-Lipska, K. and Warr, N. and Zieli{\'{n}}ska, M.},
doi = {10.1103/PhysRevC.105.024323},
issn = {2469-9985},
journal = {Phys. Rev. C},
month = {feb},
number = {2},
pages = {024323},
title = {{Coulomb excitation of $^{222}$Rn}},
url = {https://link.aps.org/doi/10.1103/PhysRevC.105.024323},
volume = {105},
year = {2022}
}

@article{Regis-GCD,
title = {Reduced γ–γ time walk to below 50 ps using the multiplexed-start and multiplexed-stop fast-timing technique with LaBr3(Ce) detectors},
journal = {Nuclear Instruments and Methods in Physics Research Section A: Accelerators, Spectrometers, Detectors and Associated Equipment},
volume = {823},
pages = {72-82},
year = {2016},
issn = {0168-9002},
doi = {https://doi.org/10.1016/j.nima.2016.04.010},
url = {https://www.sciencedirect.com/science/article/pii/S016890021630170X},
author = {J.-M. Régis and N. Saed-Samii and M. Rudigier and S. Ansari and M. Dannhoff and A. Esmaylzadeh and C. Fransen and R.-B. Gerst and J. Jolie and V. Karayonchev and C. Müller-Gatermann and S. Stegemann},
}

@article{Pascu-150Gd,
  title = {Increasing Octupole Collectivity across the $Z=64$ Isotopic Chain: B(E3) Values in $^{150}\mathrm{Gd}$},
  author = {Pascu, S. and Y\"uksel, E. and Abhishek and Stevenson, P. and Bhat, G. H. and Mao, R. N. and Nomura, K. and Costache, C. and Li, Z. P. and M\ifmmode \u{a}\else \u{a}\fi{}rginean, N. and Mihai, C. and Naz, T. and Podoly\'ak, Zs. and Regan, P. H. and Turturic\ifmmode \u{a}\else \u{a}\fi{}, A. E. and Borcea, R. and Boromiza, M. and Bucurescu, D. and C\ifmmode \u{a}\else \u{a}\fi{}linescu, S. and Clisu, C. and Coman, A. and Dinescu, I. and Doshi, S. and Filipescu, D. and Florea, N. M. and Gandhi, A. and Gheorghe, I. and Ionescu, A. and Lic\ifmmode \u{a}\else \u{a}\fi{}, R. and M\ifmmode \u{a}\else \u{a}\fi{}rginean, R. and Mihai, R. E. and Mitu, A. and Nazir, N. and Negret, A. and Ni\ifmmode \mbox{\c{t}}\else \c{t}\fi{}\ifmmode \u{a}\else \u{a}\fi{}, C. R. and O'Sullivan, E. B. and Petrone, C. and Poulton, S. E. and Sheikh, J. A. and Singh, H. K. and Stan, L. and Toma, S. and Turturic\ifmmode \u{a}\else \u{a}\fi{}, G. and Ujeniuc, S.},
  journal = {Phys. Rev. Lett.},
  volume = {134},
  issue = {9},
  pages = {092501},
  numpages = {7},
  year = {2025},
  month = {Mar},
  publisher = {American Physical Society},
  doi = {10.1103/PhysRevLett.134.092501},
  url = {https://link.aps.org/doi/10.1103/PhysRevLett.134.092501}
}

@article{KIBEDI_BE3,
title = {REDUCED ELECTRIC-OCTUPOLE TRANSITION PROBABILITIES, B(E3;01+→31−)—AN UPDATE},
journal = {Atomic Data and Nuclear Data Tables},
volume = {80},
number = {1},
pages = {35-82},
year = {2002},
issn = {0092-640X},
doi = {https://doi.org/10.1006/adnd.2001.0871},
url = {https://www.sciencedirect.com/science/article/pii/S0092640X0190871X},
author = {T Kibedi and R.H Spear}
}

@article{BRICC,
title = {Evaluation of theoretical conversion coefficients using BrIcc},
journal = {Nuclear Instruments and Methods in Physics Research Section A: Accelerators, Spectrometers, Detectors and Associated Equipment},
volume = {589},
number = {2},
pages = {202-229},
year = {2008},
issn = {0168-9002},
doi = {https://doi.org/10.1016/j.nima.2008.02.051},
url = {https://www.sciencedirect.com/science/article/pii/S0168900208002520},
author = {T. Kibédi and T.W. Burrows and M.B. Trzhaskovskaya and P.M. Davidson and C.W. Nestor}
}

@article{ShellModel-N82,
  title = {Shell model description of the $N=82$ isotonic chain with a new effective interaction},
  author = {Yu, Y. X. and Chen, Q. Y. and Qi, Chong and Fu, G. J.},
  journal = {Phys. Rev. C},
  volume = {113},
  issue = {2},
  pages = {024318},
  numpages = {16},
  year = {2026},
  month = {Feb},
  publisher = {American Physical Society},
  doi = {10.1103/shhh-41cg},
  url = {https://link.aps.org/doi/10.1103/shhh-41cg}
}

@article{DataSheetsA148,
title = {Nuclear Data Sheets for A=148},
journal = {Nuclear Data Sheets},
volume = {208},
pages = {1-396},
year = {2026},
issn = {0090-3752},
doi = {https://doi.org/10.1016/j.nds.2026.01.001},
url = {https://www.sciencedirect.com/science/article/pii/S0090375226000013},
author = {N. Nica}
}

@article{ShellGap,
title = {Extended shell model calculation for even N = 82 isotones with a realistic effective interaction},
journal = {Nuclear Physics A},
volume = {618},
number = {1},
pages = {107-125},
year = {1997},
issn = {0375-9474},
doi = {https://doi.org/10.1016/S0375-9474(97)00123-1},
url = {https://www.sciencedirect.com/science/article/pii/S0375947497001231},
author = {A. Holt and T. Engeland and E. Osnes and M. Hjorth-Jensen and J. Suhonen}
}

@article{132Sn-CoulEx,
  title = {Enhanced Quadrupole and Octupole Strength in Doubly Magic $^{132}\mathrm{Sn}$},
  author = {Rosiak, D. and Seidlitz, M. and Reiter, P. and Na\"{\i}dja, H. and Tsunoda, Y. and Togashi, T. and Nowacki, F. and Otsuka, T. and Col\`o, G. and Arnswald, K. and Berry, T. and Blazhev, A. and Borge, M. J. G. and Cederk\"all, J. and Cox, D. M. and De Witte, H. and Gaffney, L. P. and Henrich, C. and Hirsch, R. and Huyse, M. and Illana, A. and Johnston, K. and Kaya, L. and Kr\"oll, Th. and Benito, M. L. Lozano and Ojala, J. and Pakarinen, J. and Queiser, M. and Rainovski, G. and Rodriguez, J. A. and Siebeck, B. and Siesling, E. and Sn\"all, J. and Van Duppen, P. and Vogt, A. and von Schmid, M. and Warr, N. and Wenander, F. and Zell, K. O.},
  collaboration = {MINIBALL and HIE-ISOLDE Collaborations},
  journal = {Phys. Rev. Lett.},
  volume = {121},
  issue = {25},
  pages = {252501},
  numpages = {6},
  year = {2018},
  month = {Dec},
  publisher = {American Physical Society},
  doi = {10.1103/PhysRevLett.121.252501},
  url = {https://link.aps.org/doi/10.1103/PhysRevLett.121.252501}
}

@article{Scheck_2025,
doi = {10.1088/1361-6471/ade1f1},
url = {https://doi.org/10.1088/1361-6471/ade1f1},
year = {2025},
month = {jul},
publisher = {IOP Publishing},
volume = {52},
number = {6},
pages = {065105},
author = {Scheck, M and Chapman, R and Mashtakov, K and Meeten, R and Sassarini, P L and Spagnoletti, P},
title = {Seniority-two valence-shell building blocks of the octupole phonon},
journal = {Journal of Physics G: Nuclear and Particle Physics}
}

@article{Chupp2019,
archivePrefix = {arXiv},
arxivId = {1710.02504},
author = {Chupp, T. E. and Fierlinger, Peter and Ramsey-Musolf, M. J. and Singh, J. T.},
doi = {10.1103/RevModPhys.91.015001},
eprint = {1710.02504},
issn = {0034-6861},
journal = {Rev. Mod. Phys.},
month = {jan},
number = {1},
pages = {015001},
title = {{Electric dipole moments of atoms, molecules, nuclei, and particles}},
url = {http://arxiv.org/abs/1710.02504 https://link.aps.org/doi/10.1103/RevModPhys.91.015001},
volume = {91},
year = {2019}
}

@article{Dalton2023,
archivePrefix = {arXiv},
arxivId = {2302.00214},
author = {Dalton, F. and Flambaum, V. V. and Mansour, A. J.},
doi = {10.1103/PhysRevC.107.035502},
eprint = {2302.00214},
issn = {24699993},
journal = {Phys. Rev. C},
number = {3},
publisher = {American Physical Society},
title = {{Enhanced Schiff and magnetic quadrupole moments in deformed nuclei and their connection to the search for axion dark matter}},
volume = {107},
year = {2023}
}

@article{BMF1,
  title = {Spectroscopy of reflection-asymmetric nuclei with relativistic energy density functionals},
  author = {Xia, S. Y. and Tao, H. and Lu, Y. and Li, Z. P. and Nik\ifmmode \check{s}\else \v{s}\fi{}i\ifmmode \acute{c}\else \'{c}\fi{}, T. and Vretenar, D.},
  journal = {Phys. Rev. C},
  volume = {96},
  issue = {5},
  pages = {054303},
  numpages = {14},
  year = {2017},
  month = {Nov},
  publisher = {American Physical Society},
  doi = {10.1103/PhysRevC.96.054303},
  url = {https://link.aps.org/doi/10.1103/PhysRevC.96.054303}
}

@article{BMF2,
  title = {Octupole deformation properties of the Barcelona-Catania-Paris energy density functionals},
  author = {Robledo, L. M. and Baldo, M. and Schuck, P. and Vi\~nas, X.},
  journal = {Phys. Rev. C},
  volume = {81},
  issue = {3},
  pages = {034315},
  numpages = {13},
  year = {2010},
  month = {Mar},
  publisher = {American Physical Society},
  doi = {10.1103/PhysRevC.81.034315},
  url = {https://link.aps.org/doi/10.1103/PhysRevC.81.034315}
}

@article{BMF3,
  title = {Octupole correlations in the $^{144}\mathrm{Ba}$ nucleus described with symmetry-conserving configuration-mixing calculations},
  author = {Bernard, R\'emi N. and Robledo, Luis M. and Rodr\'{\i}guez, Tom\'as R.},
  journal = {Phys. Rev. C},
  volume = {93},
  issue = {6},
  pages = {061302},
  numpages = {6},
  year = {2016},
  month = {Jun},
  publisher = {American Physical Society},
  doi = {10.1103/PhysRevC.93.061302},
  url = {https://link.aps.org/doi/10.1103/PhysRevC.93.061302}
}

@article{BMF4,
  title = {Signatures of octupole correlations in neutron-rich odd-mass barium isotopes},
  author = {Nomura, K. and Nik\ifmmode \check{s}\else \v{s}\fi{}i\ifmmode \acute{c}\else \'{c}\fi{}, T. and Vretenar, D.},
  journal = {Phys. Rev. C},
  volume = {97},
  issue = {2},
  pages = {024317},
  numpages = {10},
  year = {2018},
  month = {Feb},
  publisher = {American Physical Society},
  doi = {10.1103/PhysRevC.97.024317},
  url = {https://link.aps.org/doi/10.1103/PhysRevC.97.024317}
}

@article{EDM-Candidates,
  title = {Correlating Schiff Moments in the Light Actinides with Octupole Moments},
  author = {Dobaczewski, Jacek and Engel, Jonathan and Kortelainen, Markus and Becker, Pierre},
  journal = {Phys. Rev. Lett.},
  volume = {121},
  issue = {23},
  pages = {232501},
  numpages = {6},
  year = {2018},
  month = {Dec},
  publisher = {American Physical Society},
  doi = {10.1103/PhysRevLett.121.232501},
  url = {https://link.aps.org/doi/10.1103/PhysRevLett.121.232501}
}

@article{quasi-magic-146Gd-1,
  title = {Proton Single-Particle States above $Z=64$},
  author = {Nagai, Y. and Styczen, J. and Piiparinen, M. and Kleinheinz, P. and Bazzacco, D. and Brentano, P. V. and Zell, K. O. and Blomqvist, J.},
  journal = {Phys. Rev. Lett.},
  volume = {47},
  issue = {18},
  pages = {1259--1262},
  numpages = {0},
  year = {1981},
  month = {Nov},
  publisher = {American Physical Society},
  doi = {10.1103/PhysRevLett.47.1259},
  url = {https://link.aps.org/doi/10.1103/PhysRevLett.47.1259}
}

@article{quasi-magic-146Gd-2,
title = {Non-local mean field effect on nuclei near Z=64 sub-shell},
journal = {Physics Letters B},
volume = {680},
number = {5},
pages = {428-431},
year = {2009},
issn = {0370-2693},
doi = {https://doi.org/10.1016/j.physletb.2009.09.034},
url = {https://www.sciencedirect.com/science/article/pii/S0370269309011034},
author = {Wen Hui Long and Takashi Nakatsukasa and Hiroyuki Sagawa and Jie Meng and Hitoshi Nakada and Ying Zhang}
}

@article{Isacker-208Pb,
  title = {Shell-model study of octupole collectivity near $^{208}\mathrm{Pb}$},
  author = {Van Isacker, P. and Rejmund, M.},
  journal = {Phys. Rev. Res.},
  volume = {4},
  issue = {2},
  pages = {L022031},
  numpages = {5},
  year = {2022},
  month = {May},
  publisher = {American Physical Society},
  doi = {10.1103/PhysRevResearch.4.L022031},
  url = {https://link.aps.org/doi/10.1103/PhysRevResearch.4.L022031}
}

@article{E1-Hindered,
title = {Octupole softness and electric-dipole transitions in yrast spectroscopy},
journal = {Nuclear Physics A},
volume = {557},
pages = {515-529},
year = {1993},
issn = {0375-9474},
doi = {https://doi.org/10.1016/0375-9474(93)90566-G},
url = {https://www.sciencedirect.com/science/article/pii/037594749390566G},
author = {Ikuko Hamamoto}
}

@article{ANDREJTSCHEFF2001239,
title = {New evidence for the E1 core polarization in spherical nuclei},
journal = {Physics Letters B},
volume = {506},
number = {3},
pages = {239-246},
year = {2001},
issn = {0370-2693},
doi = {https://doi.org/10.1016/S0370-2693(01)00340-9},
url = {https://www.sciencedirect.com/science/article/pii/S0370269301003409},
author = {W. Andrejtscheff and C. Kohstall and P. {von Brentano} and C. Fransen and U. Kneissl and N. Pietralla and H.H. Pitz}
}

@article{BROWN2014115,
title = {The Shell-Model Code NuShellX@MSU},
journal = {Nuclear Data Sheets},
volume = {120},
pages = {115-118},
year = {2014},
issn = {0090-3752},
doi = {https://doi.org/10.1016/j.nds.2014.07.022},
url = {https://www.sciencedirect.com/science/article/pii/S0090375214004748},
author = {B.A. Brown and W.D.M. Rae}
}

@article{PhysRevC.45.1720,
  title = {Construction of shell-model interactions for Z\ensuremath{\gtrsim}50, N\ensuremath{\gtrsim}82 nuclei: Predictions for A=133--134 ${\mathrm{\ensuremath{\beta}}}^{\mathrm{\ensuremath{-}}}$ decays},
  author = {Chou, W.-T. and Warburton, E. K.},
  journal = {Phys. Rev. C},
  volume = {45},
  issue = {4},
  pages = {1720--1729},
  numpages = {0},
  year = {1992},
  month = {Apr},
  publisher = {American Physical Society},
  doi = {10.1103/PhysRevC.45.1720},
  url = {https://link.aps.org/doi/10.1103/PhysRevC.45.1720}
}

@article{PhysRevC.80.044320,
  title = {Shell-model study of the $N=82$ isotonic chain with a realistic effective Hamiltonian},
  author = {Coraggio, L. and Covello, A. and Gargano, A. and Itaco, N. and Kuo, T. T. S.},
  journal = {Phys. Rev. C},
  volume = {80},
  issue = {4},
  pages = {044320},
  numpages = {8},
  year = {2009},
  month = {Oct},
  publisher = {American Physical Society},
  doi = {10.1103/PhysRevC.80.044320},
  url = {https://link.aps.org/doi/10.1103/PhysRevC.80.044320}
}

@article{Bk-Corr-1,
  title = {Collectivity of the 2p-2h proton intruder band of $^{116}\mathrm{Sn}$},
  author = {Petrache, C. M. and R\'egis, J.-M. and Andreoiu, C. and Spieker, M. and Michelagnoli, C. and Garrett, P. E. and Astier, A. and Dupont, E. and Garcia, F. and Guo, S. and H\"afner, G. and Jolie, J. and Kandzia, F. and Karayonchev, V. and Kim, Y.-H. and Knafla, L. and K\"oster, U. and Lv, B. F. and Marginean, N. and Mihai, C. and Mutti, P. and Ortner, K. and Porzio, C. and Prill, S. and Saed-Samii, N. and Urban, W. and Vanhoy, J. R. and Whitmore, K. and Wisniewski, J. and Yates, S. W.},
  journal = {Phys. Rev. C},
  volume = {99},
  issue = {2},
  pages = {024303},
  numpages = {10},
  year = {2019},
  month = {Feb},
  publisher = {American Physical Society},
  doi = {10.1103/PhysRevC.99.024303},
  url = {https://link.aps.org/doi/10.1103/PhysRevC.99.024303}
}

@article{Bk-Corr-2,
  title = {Abrupt shape transition at neutron number $N=60$: $B(E2)$ values in $^{94,96,98}\mathbf{Sr}$ from fast $\ensuremath{\gamma}\ensuremath{-}\ensuremath{\gamma}$ timing},
  author = {R\'egis, J.-M. and Jolie, J. and Saed-Samii, N. and Warr, N. and Pfeiffer, M. and Blanc, A. and Jentschel, M. and K\"oster, U. and Mutti, P. and Soldner, T. and Simpson, G. S. and Drouet, F. and Vancraeyenest, A. and de France, G. and Cl\'ement, E. and Stezowski, O. and Ur, C. A. and Urban, W. and Regan, P. H. and Podoly\'ak, Zs. and Larijani, C. and Townsley, C. and Carroll, R. and Wilson, E. and Fraile, L. M. and Mach, H. and Paziy, V. and Olaizola, B. and Vedia, V. and Bruce, A. M. and Roberts, O. J. and Smith, J. F. and Scheck, M. and Kr\"oll, T. and Hartig, A.-L. and Ignatov, A. and Ilieva, S. and Lalkovski, S. and Korten, W. and M\ifmmode \u{a}\else \u{a}\fi{}rginean, N. and Otsuka, T. and Shimizu, N. and Togashi, T. and Tsunoda, Y.},
  journal = {Phys. Rev. C},
  volume = {95},
  issue = {5},
  pages = {054319},
  numpages = {7},
  year = {2017},
  month = {May},
  publisher = {American Physical Society},
  doi = {10.1103/PhysRevC.95.054319},
  url = {https://link.aps.org/doi/10.1103/PhysRevC.95.054319}
}

@article{Bk-Corr-3,
  title = {Experimental study of the lifetime and phase transition in neutron-rich $^{98,100,102}\mathrm{Zr}$},
  author = {Ansari, S. and R\'egis, J.-M. and Jolie, J. and Saed-Samii, N. and Warr, N. and Korten, W. and Zieli\ifmmode \acute{n}\else \'{n}\fi{}ska, M. and Salsac, M.-D. and Blanc, A. and Jentschel, M. and K\"oster, U. and Mutti, P. and Soldner, T. and Simpson, G. S. and Drouet, F. and Vancraeyenest, A. and de France, G. and Cl\'ement, E. and Stezowski, O. and Ur, C. A. and Urban, W. and Regan, P. H. and Podoly\'ak, Zs. and Larijani, C. and Townsley, C. and Carroll, R. and Wilson, E. and Mach, H. and Fraile, L. M. and Paziy, V. and Olaizola, B. and Vedia, V. and Bruce, A. M. and Roberts, O. J. and Smith, J. F. and Scheck, M. and Kr\"oll, T. and Hartig, A.-L. and Ignatov, A. and Ilieva, S. and Lalkovski, S. and M\ifmmode \u{a}\else \u{a}\fi{}rginean, N. and Otsuka, T. and Shimizu, N. and Togashi, T. and Tsunoda, Y.},
  journal = {Phys. Rev. C},
  volume = {96},
  issue = {5},
  pages = {054323},
  numpages = {10},
  year = {2017},
  month = {Nov},
  publisher = {American Physical Society},
  doi = {10.1103/PhysRevC.96.054323},
  url = {https://link.aps.org/doi/10.1103/PhysRevC.96.054323}
}

@article{Bk-Corr-4,
  title = {Lifetime determination in $^{190,192,194,196}\mathrm{Hg}$ via $\ensuremath{\gamma}\ensuremath{-}\ensuremath{\gamma}$ fast-timing spectroscopy},
  author = {Esmaylzadeh, A. and Gerhard, L. M. and Karayonchev, V. and R\'egis, J.-M. and Jolie, J. and Bast, M. and Blazhev, A. and Braunroth, T. and Dannhoff, M. and Dunkel, F. and Fransen, C. and H\"afner, G. and Knafla, L. and Ley, M. and M\"uller-Gatermann, C. and Schomacker, K. and Warr, N. and Zell, K.-O.},
  journal = {Phys. Rev. C},
  volume = {98},
  issue = {1},
  pages = {014313},
  numpages = {11},
  year = {2018},
  month = {Jul},
  publisher = {American Physical Society},
  doi = {10.1103/PhysRevC.98.014313},
  url = {https://link.aps.org/doi/10.1103/PhysRevC.98.014313}
}

@article{Bruno-Griffin-fast-timing,
  title = {Shape coexistence in the neutron-deficient lead region: A systematic study of lifetimes in the even-even $^{188--200}\mathrm{Hg}$ with the GRIFFIN spectrometer at TRIUMF},
  author = {Olaizola, B. and Garnsworthy, A. B. and Ali, F. A. and Andreoiu, C. and Ball, G. C. and Bernier, N. and Bidaman, H. and Bildstein, V. and Bowry, M. and Caballero-Folch, R. and Dillmann, I. and Hackman, G. and Garrett, P. E. and Jigmeddorj, B. and Kilic, A. I. and MacLean, A. D. and Patel, H. P. and Saito, Y. and Smallcombe, J. and Svensson, C. E. and Turko, J. and Whitmore, K. and Zidar, T.},
  journal = {Phys. Rev. C},
  volume = {100},
  issue = {2},
  pages = {024301},
  numpages = {12},
  year = {2019},
  month = {Aug},
  publisher = {American Physical Society},
  doi = {10.1103/PhysRevC.100.024301},
  url = {https://link.aps.org/doi/10.1103/PhysRevC.100.024301}
}

@article{Schiff-Moment-Sushkov,
  title = {Schiff moments of deformed nuclei},
  author = {Sushkov, O. P.},
  journal = {Phys. Rev. C},
  volume = {110},
  issue = {1},
  pages = {015501},
  numpages = {9},
  year = {2024},
  month = {Jul},
  publisher = {American Physical Society},
  doi = {10.1103/PhysRevC.110.015501},
  url = {https://link.aps.org/doi/10.1103/PhysRevC.110.015501}
}

@article{Schiff-Moment-Flambaum,
  title = {Enhanced nuclear Schiff moment in stable and metastable nuclei},
  author = {Flambaum, V. V. and Feldmeier, H.},
  journal = {Phys. Rev. C},
  volume = {101},
  issue = {1},
  pages = {015502},
  numpages = {11},
  year = {2020},
  month = {Jan},
  publisher = {American Physical Society},
  doi = {10.1103/PhysRevC.101.015502},
  url = {https://link.aps.org/doi/10.1103/PhysRevC.101.015502}
}

@article{shape-phase-transition-casten-prl,
  title = {Empirical Realization of a Critical Point Description in Atomic Nuclei},
  author = {Casten, R. F. and Zamfir, N. V.},
  journal = {Phys. Rev. Lett.},
  volume = {87},
  issue = {5},
  pages = {052503},
  numpages = {4},
  year = {2001},
  month = {Jul},
  publisher = {American Physical Society},
  doi = {10.1103/PhysRevLett.87.052503},
  url = {https://link.aps.org/doi/10.1103/PhysRevLett.87.052503}
}

@article{Butler-framentaion-E3-strength,
author = {Butler, P. A.},
title = {Pear-shaped atomic nuclei},
journal = {Proceedings of the Royal Society A: Mathematical, Physical and Engineering Sciences},
volume = {476},
number = {2239},
pages = {20200202},
year = {2020},
month = {07},
issn = {1364-5021},
doi = {10.1098/rspa.2020.0202},
url = {https://doi.org/10.1098/rspa.2020.0202},

}

@Article{Casten2006,
author={Casten, R. F.},
title={Shape phase transitions and critical-point phenomena in atomic nuclei},
journal={Nature Physics},
year={2006},
month={Dec},
day={01},
volume={2},
number={12},
pages={811-820},
issn={1745-2481},
doi={10.1038/nphys451},
url={https://doi.org/10.1038/nphys451}
}

@article{Reinhard1995SkI,
  author  = {Reinhard, P.-G. and Flocard, H.},
  title   = {Nuclear effective forces and isotope shifts},
  journal = {Nuclear Physics A},
  volume  = {584},
  pages   = {467--488},
  year    = {1995},
  doi     = {10.1016/0375-9474(94)00770-N}
}

@article{COLO2013142,
title = {Self-consistent RPA calculations with Skyrme-type interactions: The skyrme_rpa program},
journal = {Computer Physics Communications},
volume = {184},
number = {1},
pages = {142-161},
year = {2013},
issn = {0010-4655},
doi = {https://doi.org/10.1016/j.cpc.2012.07.016},
url = {https://www.sciencedirect.com/science/article/pii/S0010465512002627},
author = {Gianluca Colò and Ligang Cao and Nguyen {Van Giai} and Luigi Capelli}
}

@misc{colo2021user,
      title={User guide for the hfbcs-qrpa(v1) code}, 
      author={Gianluca Colò and Xavier Roca-Maza},
      year={2021},
      eprint={2102.06562},
      archivePrefix={arXiv},
      primaryClass={nucl-th}
}

@article{QRPA_SkM,
  author  = {J. Bartel and P. Quentin and M. Brack and C. Guet and H.-B. H{\aa}kansson},
  title   = {Towards a better parametrisation of Skyrme-like effective forces: A critical study of the SkM force},
  journal = {Nuclear Physics A},
  volume  = {386},
  pages   = {79--100},
  year    = {1982},
  doi     = {10.1016/0375-9474(82)90403-1}
}

@article{DataSheetsA146,
title = {Nuclear Data Sheets for A = 146},
journal = {Nuclear Data Sheets},
volume = {136},
pages = {163-452},
year = {2016},
issn = {0090-3752},
doi = {https://doi.org/10.1016/j.nds.2016.08.002},
url = {https://www.sciencedirect.com/science/article/pii/S0090375216300242},
author = {Yu. Khazov and A. Rodionov and G. Shulyak}
}

@article{BE2_comp,
title = {Tables of E2 transition probabilities from the first 2+ states in even–even nuclei},
journal = {Atomic Data and Nuclear Data Tables},
volume = {107},
pages = {1-139},
year = {2016},
issn = {0092-640X},
doi = {https://doi.org/10.1016/j.adt.2015.10.001},
url = {https://www.sciencedirect.com/science/article/pii/S0092640X15000406},
author = {B. Pritychenko and M. Birch and B. Singh and M. Horoi}
}

@article{PhysRevLett.41.289,
  title = {Lowest ${2}^{+}$ State in $_{64}^{146}\mathrm{Gd}_{82}$ and the Energy Gap at $Z=64$},
  author = {Ogawa, M. and Broda, R. and Zell, K. and Daly, P. J. and Kleinheinz, P.},
  journal = {Phys. Rev. Lett.},
  volume = {41},
  issue = {5},
  pages = {289--292},
  numpages = {0},
  year = {1978},
  month = {Jul},
  publisher = {American Physical Society},
  doi = {10.1103/PhysRevLett.41.289},
  url = {https://link.aps.org/doi/10.1103/PhysRevLett.41.289}
}

@article{PhysRevLett.47.1433,
  title = {Relation between the $Z=64$ Shell Closure and the Onset of Deformation at $N=88\ensuremath{-}90$},
  author = {Casten, R. F. and Warner, D. D. and Brenner, D. S. and Gill, R. L.},
  journal = {Phys. Rev. Lett.},
  volume = {47},
  issue = {20},
  pages = {1433--1436},
  numpages = {0},
  year = {1981},
  month = {Nov},
  publisher = {American Physical Society},
  doi = {10.1103/PhysRevLett.47.1433},
  url = {https://link.aps.org/doi/10.1103/PhysRevLett.47.1433}
}
\end{document}